\documentclass[aps,pra,reprint,superscriptaddress,showpacs]{revtex4-1}

\usepackage{hyperref}
\usepackage{amsfonts}
\usepackage{amsmath}
\usepackage{amssymb}
\usepackage{latexsym}
\usepackage{graphicx}
\usepackage{color}

\usepackage{soul}

\newcommand{\be}{\begin{equation}}
\newcommand{\ee}{\end{equation}}
\newcommand{\bea}{\begin{eqnarray}}
\newcommand{\eea}{\end{eqnarray}}

\newcommand{\mralpha}{{{}_{\text{o}}\alpha}}
\newcommand{\mrbeta}{{{}_{\text{o}}\beta}}

\newcommand{\mrhatalpha}{{\hat\alpha}}
\newcommand{\mrhatbeta}{{\hat\beta}}

\newcommand{\Ahat}{{\hat A}}
\newcommand{\Bhat}{{\hat B}}

\DeclareMathOperator{\Imagpart}{Im}

\DeclareMathOperator{\diag}{diag}

\date{Revised June 2013. 
Published in 
\href{http://stacks.iop.org/1367-2630/15/073052}{New J. Phys.\ 
{\bf 15} (2013) 073052}.} 

\begin{document}

\title{Mode-mixing quantum gates and entanglement without particle creation
in periodically accelerated cavities}

\author{David Edward Bruschi}
\affiliation{School of Mathematical Sciences, 
University of Nottingham, 
Nottingham NG7 2RD, 
United Kingdom}
\affiliation{School of Electronic and Electrical Engineering, 
University of Leeds, 
Leeds LS2 9JT, 
United Kingdom}
\author{Jorma Louko}
\affiliation{School of Mathematical Sciences, 
University of Nottingham, 
Nottingham NG7 2RD, 
United Kingdom}
\affiliation{Kavli Institute for Theoretical Physics, 
University of California,
Santa Barbara, CA 93106-4030, USA}
\author{Daniele Faccio}
\affiliation{School of Engineering and Physical Sciences, 
David Brewster Building, 
Heriot-Watt University, SUPA, Edinburgh
EH14 4AS, United Kingdom}
\author{Ivette Fuentes} 
\thanks{Previously known as Fuentes-Guridi and Fuentes-Schuller.}
\affiliation{School of Mathematical Sciences, 
University of Nottingham, 
Nottingham NG7 2RD, 
United Kingdom}

\begin{abstract}
We show that mode-mixing quantum gates can be produced 
by non-uniform relativistic acceleration. 
Periodic motion in cavities exhibits a series of resonant 
conditions producing entangling quantum gates between 
different frequency modes. The resonant condition associated 
with particle creation is the main feature of the dynamical 
Casimir effect which has been recently demonstrated in 
superconducting circuits. We show that a second resonance, 
which has attracted less attention since it implies 
negligible particle production, produces a beam splitting 
quantum gate leading to a resonant enhancement of entanglement 
which can be used as the first evidence of acceleration effects 
in mechanical oscillators. We propose a desktop experiment 
where the frequencies associated with this second 
resonance can be produced mechanically.
\end{abstract}


\pacs{04.62.+v, 42.50.Xa, 42.50.Dv, 42.50.Pq} 





\maketitle

\section{Introduction} 

In relativistic quantum field theory, the particle content of a
quantum state is affected by the evolution of the spacetime, including
the motion of any boundaries. Further, the very notion of a
``particle'' depends on the motion of an observer. In flat spacetime,
celebrated examples are the thermality seen in Minkowski vacuum by
uniformly accelerated observers, known as the Unruh
effect~\cite{unruh,Crispino:2007eb}, and the creation of particles by
moving boundaries, known as the dynamical (or non-stationary)
Casimir effect (DCE)~\cite{Dodonov:advchemphys,Dodonov:2010zza}. In
curved spacetime, a celebrated example is the Hawking radiation
emitted by black holes~\cite{hawking}. The DCE is related to a
fundamental prediction by Fulling and Davies that a non-uniformly
accelerated mirror will excite photons out of the
vacuum~\cite{Fulling:Davies:76}. It was later realised that this
effect may be significantly enhanced if, instead of a simple mirror, a
cavity {is used} in which one or both of the mirrors are in motion
\cite{Reynaud1}. The simplest situation in which to observe the DCE is
that of a cavity oscillating sinusoidally with frequency~$\omega_c$.
The DCE is predicted to exhibit a fundamental resonance condition for
the production of quantum entangled photon-pairs,
$\omega_c=\omega_1+\omega_2$, where $\omega_{1,2}$ are the two
entangled photon frequencies~\cite{Reynaud1}. The actual number of
photons predicted for a mechanically oscillating cavity is strongly
limited ($\sim10^{-9}$ photons/second) by the maximum
achievable~$\omega_c$. For this reason a number of alternative systems
that also exhibit a periodically varying boundary of some kind have
been proposed with the aim of enhancing the DCE. Examples are
superconducting quantum interference device (SQUID) mirrors, 
Bose-Einstein condensates (producing
phonon pairs) and cavities controlled using nonlinear optics
\cite{Dodonov:advchemphys,Dodonov:2010zza,wilson-etal,Westbrook}.
Notwithstanding recent breakthroughs, the DCE remains an extremely
difficult effect to observe and study experimentally.

In this paper we consider the  general case of a rigid 
cavity undergoing an arbitrary (mechanically induced) acceleration. 
In the specific cases of a linear sinusoidal or a uniform circular motion,
we show that a mode mixing resonance condition,
$\omega_c=|\omega_1 - \omega_2|$~\cite{Mundarain:1998kz},
for which no photons are generated, 
can be brought significantly below the 
DCE photon generation resonance condition,
to apparently experimentally accessible frequencies. 
We show how this low-frequency resonance leads to the generation of
entanglement between existing and previously non-entangled cavity
modes. The oscillating cavity can be shown to behave like a
generalised beam-splitter, thus performing an essential quantum gate
functionality. 
This demonstrates that relativistic effects, 
in this case non-uniform relativistic acceleration, can be
exploited for quantum information. 
There are many proposals to generate gates in non-relativistic quantum
information. Our scheme pioneers on how to implement 
quantum gates in relativistic quantum information.
We then discuss the possibility of performing actual
experiments with mechanically oscillating optical cavities.

\section{Cavity in $(1+1)$ dimensions}

\subsection{Preliminaries}

We first consider the simplified case of a cavity in
$(1+1)$-dimensional Minkowski spacetime.  The cavity is assumed
mechanically rigid, maintaintaining constant length $L$ in its
instantaneous rest frame.  The proper acceleration at the centre of
the cavity is denoted by $a(\tau)$, where $\tau$ is the proper time.
To maintain rigidity, the acceleration must be bounded by
$|a(\tau)|L/c^2 < 2$~\cite{Bruschi:2011ug}.  From now on we set
$c=\hbar=1$.

The cavity contains a real scalar field $\phi$ of mass $\mu_0\ge0$,
with Dirichlet boundary conditions. We assume that the cavity is
initially inertial, and we denote by $u_n$, $n=1,2,\ldots$, a standard
basis of cavity field modes that are of positive frequency $\omega_n =
\sqrt{\mu_0^2 + (\pi n/L)^2}$ with respect to the cavity's proper time
before the acceleration. We also assume that the cavity's final state
is inertial, and we denote by $\tilde u_n$, $n=1,2,\ldots$, a standard
basis of cavity field modes that are of positive frequency $\omega_n$
with respect to the cavity's proper time after the
acceleration. Because of the acceleration at intermediate times, the
two sets of modes need not coincide, but the completeness of each set
allows the sets to be related by the Bogoliubov transformation
\cite{byd,fabbri-navarro-salas}
\begin{align}
\tilde u_m = \sum_n \, \bigl( 
\alpha_{mn} u_n + \beta_{mn} u^*_n
\bigr)
\ , 
\label{eq:bogotransf-general}
\end{align}
where the star denotes complex conjugation. The Bogoliubov
coefficient matrices $\alpha$ and $\beta$ are determined by solving
the field equation in the cavity during the acceleration. 

In the initial and final inertial regions the field operator
$\phi$ has the respective expansions 
$\phi = \sum_n
\left( a_n u_n +  a_n^\dagger u^*_n\right)$ 
and 
$\phi = \sum_n
\left( \tilde a_n \tilde u_n +  {\tilde a}_n^\dagger {\tilde u}^*_n\right)$, 
where the nonvanishing commutators of the early (respectively late) 
time creation and
annihilation operators are 
$\bigl[a_n,a_m^\dagger\bigr] = \delta_{nm}$ 
$\bigl(\bigl[\tilde a_n, {\tilde a}_m^\dagger\bigr] = \delta_{nm}\bigr)$. 
The early and late time creation and annihilation operators need not
coincide, but it follows from \eqref{eq:bogotransf-general} that they 
can be expressed in terms of each other in terms of
the Bogoliubov coefficients~\cite{byd,fabbri-navarro-salas}. In particular, 
the transformation mixes creation and annihilation operators 
if and only if some of the 
$\beta$-coefficients are nonvanishing. 

Now, working in the Heisenberg picture, the quantum state of the field
in the cavity does not change in time. Howevever, given a
state~$\left|\Psi\right>$, we interpret its particle content at early
times in terms of the early time vacuum~$\left|0\right>$, which
satisfies $a_n\left|0\right>=0$, and the early time excitations
created by~$a_n^\dagger$. At late times, we similarly interpret the
particle content of $\left|\Psi\right>$ in terms of the late time
vacuum~$\left|\tilde 0\right>$, which satisfies 
$\tilde a_n\left|\tilde0\right>=0$, and the late time excitations created
by~${\tilde a}_n^\dagger$. The acceleration hence affects the particle
content of the cavity whenever the Bogoliubov transformation
\eqref{eq:bogotransf-general} differs from the identity
transformation. The $\beta$-coefficients are responsible for creation
and annihilation of particles, while the $\alpha$-coefficients are
responsible for mode mixing. In particular, the vacua $\left|0\right>$
and $\left|\tilde 0\right>$ coincide if and only if all the
$\beta$-coefficients vanish~\cite{byd,fabbri-navarro-salas}.

\subsection{Bogoliubov coefficients for general 
time-dependent acceleration}

We shall express the Bogoliubov coefficients as a time-ordered integral,
allowing both the magnitude and the time-dependence of the
acceleration to remain general within the rigidity bound
$|a(\tau)|L < 2$.

We encode $\alpha$ and $\beta$
into the matrix $U = \left( \begin{smallmatrix}
    \alpha &\beta \\
    \beta^* & \alpha^*
\end{smallmatrix}
\right)$, so that the composition of Bogoliubov
transformations amounts to matrix multiplication of the corresponding 
$U$-matrices. The 
Bogoliubov identities \cite{byd} are then encoded in the matrix equation 
$\left(\begin{smallmatrix}
1 &0 \\
0 & -1 
\end{smallmatrix}\right) = 
U \left(\begin{smallmatrix}
1 &0 \\
0 & -1 
\end{smallmatrix} \right) 
U^\dagger
$.  

When the acceleration between the initial and final 
inertial regions is uniform and lasts for proper time~$\bar\tau$, 
we have \cite{Bruschi:2011ug} 
$
U_h(\bar\tau) = K_h^{-1} \tilde Z_h(\bar\tau)  K_h$,  
where 
$K_h = 
\left(\begin{smallmatrix}
\mralpha_h &\mrbeta_h \\
\mrbeta_h & \mralpha_h 
\end{smallmatrix}\right)$, 
$\tilde Z_h(\bar\tau) = 
\left(\begin{smallmatrix}
{Z}_h(\bar\tau) &0 \\
0 & {Z}_h^{*}(\bar\tau)
\end{smallmatrix}\right)$, 
${Z}_h(\bar\tau) = 
\diag \bigl(
e^{i\Omega_1(h)\bar\tau}, e^{i\Omega_2(h)\bar\tau},
\cdots\bigr)$, 
$\Omega_n(h)$ are the angular frequencies during the acceleration, 
$\mralpha_h$ and $\mrbeta_h$ are the Bogoliubov coefficient matrices
from the initial inertial segment to the uniformly accelerated
segment, and the acceleration has been encoded in the dimensionless
parameter $h= aL$.  The field modes and the angular frequencies during
the acceleration have
elementary expressions for $\mu_0=0$ and are given in terms of
modified Bessel functions for $\mu_0>0$. The coefficients encoded in 
$\mralpha_h$ and $\mrbeta_h$ do not have elementary expressions, but
they can be written as integrals 
involving the inertial and accelerated mode functions over the
constant time surface where the acceleration begins. 

For accelerations that may vary arbitrarily between 
the initial time $\tau_0$ and final time~$\tau$, 
$U(\tau,\tau_0)$ is given by the limit of 
$U_{h_N}(\bar\tau_N) U_{h_{N-1}}(\bar\tau_{N-1})\cdots
U_{h_2}(\bar\tau_2) U_{h_1}(\bar\tau_1)$ as $N\to\infty$, 
such that $\tau-\tau_0 = \sum_{k=1}^N \bar\tau_{k}$ 
is fixed and each $\bar\tau_k\to0$. 
As an infinitesimal increase in $\tau$
amounts to multiplying $U(\tau,\tau_0)$ from the left by 
$U_h(\bar\tau)$ with infinitesimal~$\bar\tau$, 
$U(\tau,\tau_0)$ satisfies the differential equation 
\begin{align}
{\dot U}(\tau,\tau_0) = i 
K_{h(\tau)}^{-1} \tilde{\Omega}_{h(\tau)} K_{h(\tau)}
{U}(\tau,\tau_0) \,,
\label{eq:U-diffeq}
\end{align}
where 
$\tilde\Omega_{h(\tau)} 
= \left(\begin{smallmatrix}
{\Omega}_{h(\tau)} &0 \\
0 & - {\Omega}_{h(\tau)}
\end{smallmatrix}\right)$, 
${\Omega}_{h(\tau)}
= 
\diag \bigl(
\Omega_1\bigl(h(\tau)\bigr), 
\Omega_2\bigl(h(\tau)\bigr),\cdots \bigr)$, 
and the overdot denotes derivative with respect to~$\tau$. 
The solution is 
\begin{align}
{U}(\tau_f,\tau_0)  = T \exp\left(i\int_{\tau_0}^{\tau_f} 
K_{h(\tau)}^{-1} \tilde\Omega_{h(\tau)} K_{h(\tau)}
\, d\tau \right) 
\,,
\label{eq:U-solution}
\end{align}
where
$\tau_f$ denotes the moment at which the acceleration ends and 
$T$ denotes the time-ordered exponential. 

To summarise: the Bogoliubov transformation between the inertial
initial segment ending at proper time $\tau_0$ and the final inertial
segment starting at proper time $\tau_f$ is given
by~\eqref{eq:U-solution}.  $h(\tau)$~may vary arbitrarily for $\tau_0
\le \tau \le \tau_f$, within the rigidity constraint $|h(\tau)|<2$: in
particular, no small acceleration approximation has been made.  For
piecewise constant~$h(\tau)$, \eqref{eq:U-solution} reduces to a
product of the matrices $U_h(\bar\tau_k)$ from each constant $h$
segment.

A direct consequence of \eqref{eq:U-solution} is that the Bogoliubov
coefficients evolve by pure phases over any time interval in which $h$
is constant.  Particles in the cavity are hence created by
\emph{changes\/} in the acceleration, not by acceleration itself, as
can be argued on general adiabaticity grounds
\cite{schutzhold-unruh:comment-tele,unruh-private}. The cavity is in
this respect similar to a single accelerating mirror, which excites
photons from the vacuum only when its acceleration is
non-uniform~\cite{Fulling:Davies:76}.

\subsection{Small acceleration limit}

At small accelerations, $\Omega_n(h)$, $\mralpha_h$ and $\mrbeta_h$ 
have the expansions \cite{Bruschi:2011ug} 
\begin{subequations}
\label{eq:Omega-ab-expansion}
\begin{align}
\Omega_n &= \omega_n + O(h^2) \ , 
\hspace{2ex}
n=1,2,\ldots, 
\label{eq:Omega-pert}
\\
\mralpha_h &= 1 + h \mrhatalpha + O(h^2) , 
\hspace{2ex}
\mrbeta_h = h \mrhatbeta + O(h^2) , 
\label{eq:ab-expansion}
\end{align}
\end{subequations} 
where 
\begin{subequations}
\label{eq:dirichlet-bogocoeffs-linear}
\begin{align}
\mrhatalpha_{nn} &=0 
\ , 
\\
\mrhatalpha_{mn} &= 
\frac{\pi^2 m n \bigl(-1+{(-1)}^{m+n}\bigr)} 
{L^4\left(\omega_m - \omega_n\right)^3
\sqrt{\omega_m \omega_n}}
\ \ \ \hbox{for $m\ne n$}, 
\\[1ex]
\mrhatbeta_{mn} &=
\frac{\pi^2 m n \bigl(1 - {(-1)}^{m+n}\bigr)} 
{L^4\left(\omega_m + \omega_n\right)^3
\sqrt{\omega_m \omega_n}}
\ . 
\end{align}
\end{subequations}
Note that 
$\mrhatalpha_{mn}$ and $\mrhatbeta_{mn}$ 
depend on $\mu_0$ and $L$ only via the dimensionless quantity~$\mu_0 L$. 
Formulas \eqref{eq:dirichlet-bogocoeffs-linear} are obtained from
equations (7) in \cite{Bruschi:2011ug} by an elementary
rearrangement. 

We seek $U(\tau_f,\tau_0)$ {in the form} 
\begin{subequations}
\label{eq:AhatBhat-defs}
\begin{align}
\alpha &= e^{i\boldsymbol{\omega} (\tau_f-\tau_0)} \bigl(1 + \Ahat + O(h^2) \bigr),
\\
\beta &= e^{i\boldsymbol{\omega} (\tau_f-\tau_0)} \Bhat + O(h^2), 
\end{align}
\end{subequations}
where
$
\boldsymbol{\omega} = 
\diag (
\omega_1, \omega_2,\cdots )
$, $\Ahat$ and $\Bhat$ are of first order in~$h$, 
and $\tau_f$ again denotes the moment at which the acceleration ends. 
Using \eqref{eq:U-diffeq}, \eqref{eq:Omega-ab-expansion}
and~\eqref{eq:dirichlet-bogocoeffs-linear},  
we find 
\begin{subequations}
\label{eq:ABhat-int-components}
\begin{align}
\Ahat_{mn} &= 
i (\omega_m-\omega_n) \mrhatalpha_{mn}
\int_{\tau_0}^{\tau_f} e^{-i({\omega}_m - {\omega}_n) (\tau-\tau_0)}
\, h(\tau) \, d\tau 
\ , 
\\
\Bhat_{mn} &=
i (\omega_m+\omega_n) \mrhatbeta_{mn}
\int_{\tau_0}^{\tau_f} e^{-i({\omega}_m + {\omega}_n) (\tau-\tau_0)} \, h(\tau) \, d\tau 
\ . 
\label{eq:Bhat-int-components}
\end{align}
\end{subequations}
To linear order in~$h$, the Bogoliubov coefficients are hence obtained 
by just Fourier transforming the acceleration.


Two comments are in order. 
First, while the perturbative solution
\eqref{eq:ABhat-int-components}
assumes the acceleration to be so small that $|h| \ll1$, 
the velocities, travel times and travel
distances remain unrestricted, 
and the solution remains valid even when the velocities are relativistic. 
Our perturbative treatment is hence complementary to the small distance
approximations often considered in the DCE
literature~\cite{Dodonov:advchemphys,Dodonov:2010zza}, while of course
overlapping in the common domain of validity.

Second, $\Ahat_{mn}$ and $\Bhat_{mn}$ scale linearly in~$h$, but their
magnitudes depend also crucially on whether $h$ changes slowly or
rapidly compared with the oscillating 
integral kernels in~\eqref{eq:ABhat-int-components}. 
In the limit of slowly-varying $h$ both $\Ahat_{mn}$ and $\Bhat_{mn}$ 
vanish, in agreement with the adiabaticity arguments 
of \cite{schutzhold-unruh:comment-tele,unruh-private}.
In the limit of piecewise constant~$h$, 
the changes in the magnitudes of 
$\Ahat_{mn}$ and $\Bhat_{mn}$ come entirely from
the discontinuous jumps in $h$ \cite{Bruschi:2011ug,Friis:2011yd,Friis:2012tb}. 
The limit of piecewise constant $h$ may 
be difficult to realise experimentally with 
a material cavity, 
and we emphasise that no such rapid changes in the acceleration are 
involved in the 
experimental scenario considered in Section~\ref{sec:experiment}. 
This limit can however be simulated by
a cavity whose walls are mechanically static 
dc SQUIDs undergoing electric
modulation~\cite{wilson-etal,Friis:2012cx}.

\subsection{Resonances}
\label{subsec:resonances}

Suppose now that $|h| \ll1$, so that the solution 
\eqref{eq:AhatBhat-defs}--\eqref{eq:ABhat-int-components} is
valid. Suppose that $h$ is sinusoidal with angular
frequency~$\omega_c$. 
For generic values of $\omega_c$ the integrals in 
\eqref{eq:ABhat-int-components} are oscillatory and have no net growth
as $\tau_f$ increases. However, when 
$\omega_c$ equals the angular frequency of an oscillating 
integral kernel in~\eqref{eq:ABhat-int-components}, 
there is a resonance and the corresponding 
Bogoliubov coefficient grows linearly in~$\tau_f$. 
These resonance conditions read 
\begin{subequations}
\begin{align}
&\Ahat_{mn}:
\hspace{1ex}
\omega_c = |\omega_m - \omega_n| 
\,,
\label{eq:Ahat-lin-resonance}
\\
&\Bhat_{mn}:
\hspace{1ex}
\omega_c = \omega_m + \omega_n 
\,, 
\label{eq:Bhat-lin-resonance}
\end{align}
\end{subequations}
where in each case $m-n$ needs to be odd 
in order for the coefficient
to be nonvanishing. 

The particle creation resonance \eqref{eq:Bhat-lin-resonance}
is well known in the DCE 
literature~\cite{Reynaud1,Dodonov:advchemphys,Dodonov:2010zza,Mundarain:1998kz,Crocce:2001zz,Crocce:2002hd,Ruser:2005xg,Yuce:2008tb,ISI:000172183100003,yuce-ozcak,sch-plunien-soff,dodonov-andreata:1999,dodo-dodo-mizrahi:2005}. 
The mode mixing resonance \eqref{eq:Ahat-lin-resonance} has been noted  
\cite{Mundarain:1998kz,Crocce:2001zz,Crocce:2002hd,Ruser:2005xg,Yuce:2008tb,ISI:000172183100003,yuce-ozcak} 
but seems to have received attention mainly in 
situations where it happens to coincide with a particle creation
resonance. 
As the case of interest in the experimental scenario of Section
\ref{sec:experiment} will be mode mixing without significant particle creation, we
recall here some relevant properties of mode mixing in quantum optics. 

Mode mixing without particle creation is known in quantum optics as a
passive transformation~\cite{simon-mukunda-dutta}, implemented
experimentally by passive optical elements such as beam splitters and
phase plates. While mode mixing is present already in classical wave optics, 
its significance in quantum optics is that the mixing 
can be harnessed to quantum information tasks. The 
entangling power of passive transformations is well 
understood: for example, the mixing generates entanglement 
from an initial Gaussian state only if this state is 
squeezed~\cite{Friis:2012tb,PhysRevLett.90.047904,Friis:2012nb,Alsing:Fuentes:2012}. 

These entanglement considerations are directly applicable to mode mixing in our cavity, 
and the entanglement can be determined experimentally by measurements on 
quanta that are allowed to
escape from the cavity~\cite{briegel:maser,lutterbach:wigner,lougovski:wigner,pielawa:generation}. 
We emphasise that while the
particle creation resonance \eqref{eq:Bhat-lin-resonance} can be used to 
implement two-mode squeezing gates
\cite{Bruschi:2012uf,Friis:2012nb,Alsing:Fuentes:2012} and other
multipartite gates~\cite{Friis:2012ki}, 
the mode-mixing resonance \eqref{eq:Ahat-lin-resonance}
can be used to implement mode-mixing gates 
even when no particle creation is present.

Specifically, the oscillating cavity can be tuned to act as a 
beam splitter --- a well-studied quantum gate in 
continuous variable systems~\cite{Weedbrook:Pirandola:2012}. 
In the following we apply our results to the special class of Gaussian states, 
characterized by positive Wigner functions, which allow elegant and powerful 
analytical results~\cite{Laurat:2005}. 
Gaussian states, including
coherent and squeezed states,
are routinely prepared in the laboratory. 
For example, two-mode squeezed states are commonly 
produced by parametric down conversion~\cite{Wu:87}. 

In order to calculate the entanglement generated by the mode mixing 
gate we have introduced, 
we adopt the covariance matrix formalism. 
We consider a family of 
harmonic oscillators 
with position and momentum operators $q_i,p_i$, 
where $i = 1,2,\dots$. 
We collect these operators in the vector 
$\mathbb{X}=(q_1,p_1,q_2,p_2,\ldots)$. 
The canonical commutation relations take the form 
$[\mathbb{X}_i,\mathbb{X}_j]=i\Omega_{ij}$, 
where the only nonvanishing components of the symplectic form 
$\mathbf{\Omega}$ are $\Omega_{2i-1,2i} = - \Omega_{2i,2i-1}= 1$. 
The covariance matrix is defined by 
\begin{align} 
\sigma_{ij}=\tfrac{1}{2}\langle\mathbb{X}_i\mathbb{X}_j+\mathbb{X}_j\mathbb{X}_i\rangle
-\langle\mathbb{X}_i\rangle\langle\mathbb{X}_j\rangle
\ .
\end{align}
This formalism is suitable for Gaussian states since all the relevant 
information about the state can be encoded in the first 
moment $\langle\mathbb{X}_i\rangle$ 
and the covariance matrix. In fact, the quantification of 
Gaussian state entanglement requires only its covariance matrix~\cite{Laurat:2005}. 

An initial pure state remains pure to first order in the acceleration~\cite{Bruschi:2012uf}. 
This implies that the reduced state $\sigma_{\text{red}}$ of two modes will depend only on the
Bogoliubov coefficients that mix these two modes. 
The contribution of coefficients mixing with other frequencies 
is negligible to linear order in the acceleration. 

From now on we specialise to two-mode Gaussian states for which 
$\sigma_{\text{red}}$ is symmetric. 
In the covariance matrix formalism, 
the entanglement for such states is fully 
quantified by the smallest symplectic eigenvalue 
of the partial transpose of the covariance matrix~\cite{Laurat:2005}. 
The partial transpose is given by 
$\tilde{\sigma}_{\text{red}}=\mathbf{P}\sigma_{\text{red}}\mathbf{P}$, 
where $\mathbf{P}=\diag(1,1,1,-1)$, and its symplectic 
eigenvalues are the eigenvalues of
$i\mathbf{\Omega}\tilde{\sigma}_{\text{red}}$.
The eigenvalue set has the form 
$\{-\tilde{\nu}_-,\tilde{\nu}_-,-\tilde{\nu}_+,\tilde{\nu}_+\}$, 
where $0\le \tilde{\nu}_- \le \tilde{\nu}_+$, 
and therefore the quantity characterising the entanglement is~$\tilde{\nu}_-$. 
Within our perturbative small acceleration expansion, 
it is shown in \cite{Friis:2012nb} that 
$\tilde{\nu}_-=1-\tilde{\nu}_-^{(1)} + O(h^2)$, 
where $\tilde{\nu}_-^{(1)}$ is linear in the acceleration. 

%

While the von Neumann entropy would be a natural measure of entanglement 
for the reduced two mode state~$\sigma_{\text{red}}$ when the initial state is pure, 
its small acceleration expansion does not take the form of a 
power series because of the logarithms involved in its 
definition~\cite{Bruschi:2012uf,Weedbrook:Pirandola:2012}.
Measures of entanglement based on the partial transpose criterion however do 
have a power series expansion in the acceleration. 
We consider the negativity~\cite{vidal-werner:computable}, 
which is in the covariance matrix formalism given by 
$\mathcal{N}=\max \left\{0, \tfrac12 ( \tilde{\nu}_-^{-1} -1) \right\}$ 
\cite{Laurat:2005}, 
and which hence has the small acceleration expansion 
$\mathcal{N}=\max \bigl\{0, \tfrac12 \tilde{\nu}_-^{(1)} \bigr\} + O(h^2)$. 

Suppose now that the state is a separable state of two modes, 
$m$ and~$n$, in which each mode is squeezed with 
the same squeezing parameter $s>0$. 
When $\Bhat_{mn}$ is negligible compared with~$\Ahat_{mn}$, it follows from the 
expression of $\tilde{\nu}_-^{(1)}$ given in \cite{Friis:2012nb} that  
the leading order contribution to the negativity 
is linear in the acceleration and given by 
\begin{align}
\mathcal{N} =  
\bigl| \Imagpart 
\bigl(
\Ahat_{mn}
\bigr)
\bigr| 
\sinh s 
\ . 
\label{eq:firstorderneg}
\end{align}

\begin{figure}
\includegraphics[scale=0.44]{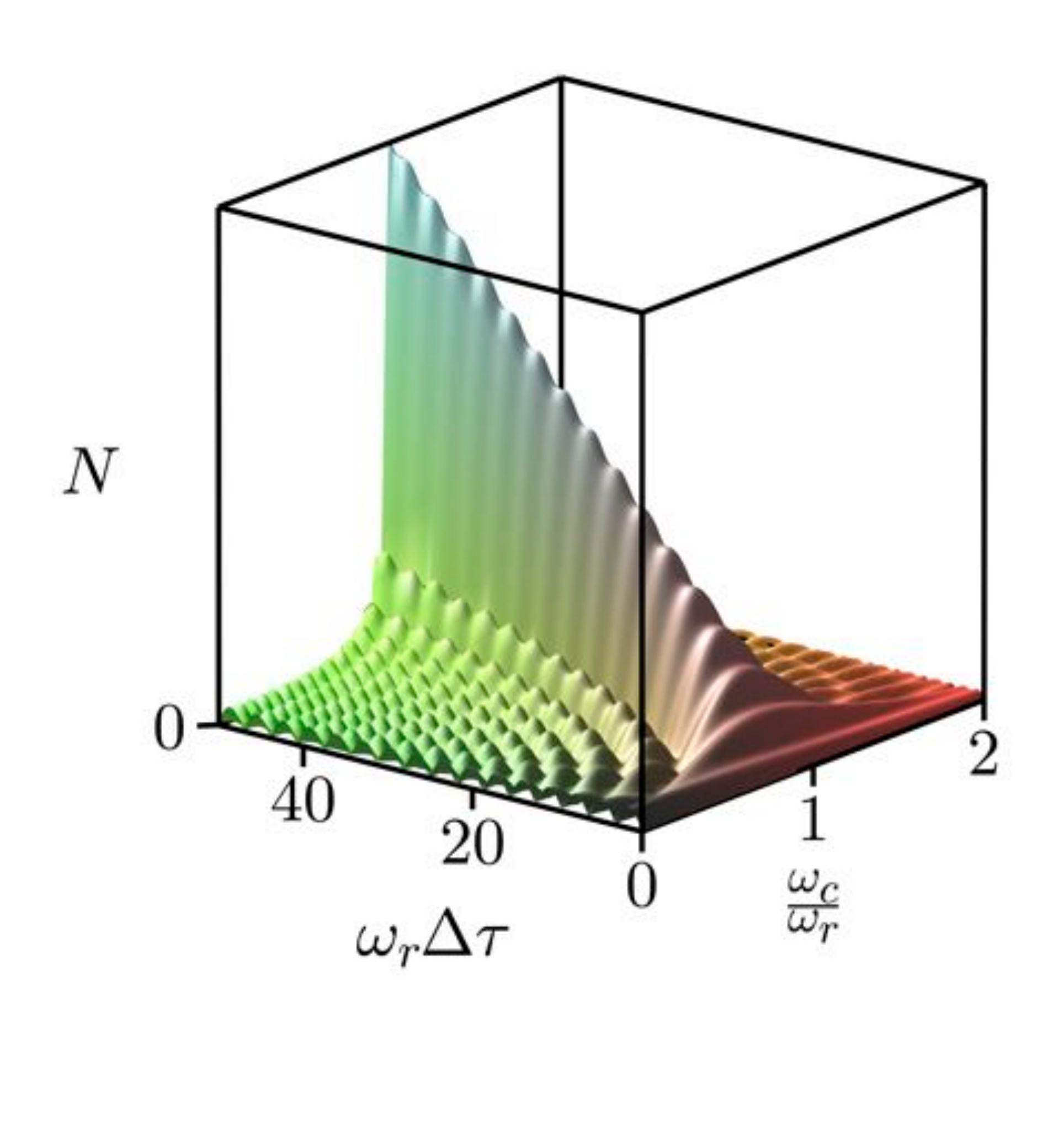} 
\caption{\label{fig:alpharesneg}The negativity $\mathcal{N}$
\eqref{eq:firstorderneg} as a function of
the cavity oscillation angular frequency $\omega_c$ and the total
oscillation time $\Delta\tau = \tau_f - \tau_0$, for fixed 
$\omega_r = |\omega_m - \omega_n|$ and squeezing parameter~$s$.}
\end{figure}

Figure \ref{fig:alpharesneg} shows a plot of the 
negativity \eqref{eq:firstorderneg} as a function of 
the total oscillation time 
$\Delta\tau = \tau_f - \tau_0$ and the acceleration 
angular frequency~$\omega_c$, 
assuming purely sinusoidal acceleration with phase chosen so that 
$h(\tau)$ is proportional to 
$\cos\bigl(\omega_c(\tau-\tau_0)\bigr)$, for fixed
$\omega_r = |\omega_m - \omega_n|$ and squeezing parameter~$s$. 
The linear growth of $\mathcal{N}$ at the resonance, 
$\omega_c = \omega_r$, is evident from the plot. 
The scale of the vertical axis depends on $m$, $n$ and $s$ 
but also on~$\mu_0L$, 
in a way that we shall address in Section~\ref{sec:experiment}.

\section{Cavity in $(3+1)$ dimensions}

Let now $\phi$ be a real scalar field of mass 
$\mu\ge0$ in a cavity in $(3+1)$-dimensional Minkowski space, 
with Dirichlet conditions. 
The inertial cavity is a rectangular parallelepiped with
fixed edge lengths $L_x$, $L_y$ and~$L_z$, 
and a standard basis of orthonormal
field modes is indexed by triples $(m,n,p)$ of positive integers, such that
the angular frequencies are 
$
\omega_{mnp} = 
\sqrt{\mu^2 
+ (\pi m / L_x)^2 
+ (\pi n / L_y)^2 
+ (\pi p / L_z)^2}
$. 

%

Acceleration in the cavity's three principal directions 
can be treated as $(1+1)$-dimensional, with the inert transverse quantum numbers
just contributing to the effective mass. 
Acceleration of unrestricted magnitude and direction would require new
input regarding how the shape of the cavity responds to such
acceleration~\cite{Epp:2008kk}. 
To \emph{linear\/} order in the acceleration, however,
boosts commute, and we can treat acceleration as a
vector superposition of accelerations in the three principal directions
in the cavity's instantaneous rest frame. Defining $\Ahat$ and $\Bhat$ as
in~\eqref{eq:AhatBhat-defs}, and denoting the acceleration three-vector in the 
cavity's instantaneous rest frame by 
$\bigl(a_x(\tau), a_y(\tau), a_z(\tau) \bigr)$, 
equations \eqref{eq:ABhat-int-components} generalise
for changes in the quantum number $m$ to 
\begin{widetext}
\begin{subequations}
\label{eq:3+1ABhat-int-components}
\begin{align}
\Ahat_{mnp,mnp} &= 0 
\ , 
\\
\Ahat_{mnp,m'np} &= 
i 
\frac{\pi^2 m m' \bigl(-1+{(-1)}^{m+m'}\bigr)} 
{L_x^3\left(\omega_{mnp} - \omega_{m'np}\right)^2
\sqrt{\omega_{mnp} \omega_{m'np}}}
\int_{\tau_0}^{\tau_f} e^{-i(\omega_{mnp} - \omega_{m'np}) (\tau-\tau_0)} \, a_x(\tau) \, d\tau 
\ \ \ \hbox{for $m\ne m'$}, 
\label{eq:3+1Ahat-int-components}
\\[1ex]
\Bhat_{mnp,m'np} &=
i 
\frac{\pi^2 m m' \bigl(1 - {(-1)}^{m+m'}\bigr)} 
{L_x^3\left(\omega_{mnp} + \omega_{m'np}\right)^2
\sqrt{\omega_{mnp} \omega_{m'np}}}
\int_{\tau_0}^{\tau_f} e^{-i(\omega_{mnp} + \omega_{m'np}) (\tau-\tau_0)} \, a_x(\tau) \, d\tau 
\ , 
\label{eq:3+1Bhat-int-components}
\end{align}
\end{subequations}
\end{widetext}
with changes in the quantum numbers $n$ and $p$ given by similar formulas 
involving respectively $a_y$ and~$a_z$. 

For sinusoidal acceleration with angular frequency~$\omega_c$, the
resonance condition of linear growth is 
\begin{subequations}
\label{eq:3+1resonance}
\begin{align}
&\Ahat_{mnp,m'n'p'}:
\hspace{1ex}
\omega_c = |\omega_{mnp} - \omega_{m'n'p'}|
\,,
\label{eq:3+1resonance-Ahat}
\\
&\Bhat_{mnp,m'n'p'}: 
\hspace{1ex}
\omega_c = \omega_{mnp} + \omega_{m'n'p'}
\,,
\label{eq:3+1resonance-Bhat}
\end{align}
\end{subequations}
where in each case only the quantum number in the 
direction of the oscillation may differ and this 
difference needs to be odd.

\section{Desktop experiment}
\label{sec:experiment}

The particle creation resonance angular frequency 
\eqref{eq:3+1resonance-Bhat} is always larger than than the 
frequencies of the individual cavity modes. 
The mode mixing resonance angular frequency
\eqref{eq:3+1resonance-Ahat} can however be lower.
In $(1+1)$ dimensions, mode mixing resonances occur 
significantly below the frequencies of the 
individual cavity modes if $\mu_0 L\gg1$, as then 
$(\Delta \omega_n)/\omega_n \approx 
\pi^2 {(\mu_0 L)}^{-2} n \Delta n \ll1$ whenever 
$n$ and $\Delta n$ are small compared with~$\mu_0 L$. 
In more than $(1+1)$ dimensions, 
a similar lowering 
can be arranged to occur 
even for a massless field by storing in the cavity quanta whose 
wave vector is highly transverse to the acceleration, 
as the transverse momentum then gives rise to a large 
effective $(1+1)$-dimensional mass. 
We now outline a $(3+1)$-dimensional experimental 
scenario that optimises this lowering 
of the mode mixing resonance. 

Setting $\mu=0$, we assume that the quanta in the cavity have 
wavelength $\lambda \ll \min(L_x,L_y,L_z)$ 
and have their momenta aligned close to the 
$z$-direction, so that 
${(2/\lambda)}^2
\approx 
{(p/L_z)}^2 \gg 
{(m/L_x)}^2 + {(n/L_y)}^2$ 
and 
$\omega_{mnp} 
\approx 
2\pi/\lambda + 
\tfrac14 \pi\lambda
\bigl[
{(m/L_x)}^2 + {(n/L_y)}^2
\bigr]
$. 
We let the cavity undergo linear or circular 
harmonic oscillation orthogonal to the $z$-direction, 
with amplitude $d_x$ ($d_y$) 
in the $x$-direction ($y$-direction). 
For motion in the $x$-direction, 
the mode mixing resonance angular frequency 
\eqref{eq:3+1resonance-Ahat} between modes $m$ and~$m'$, 
with $m-m'$ odd, is 
\begin{align}
\omega_c
& \approx 
\tfrac14 \pi \lambda L_x^{-2} \, \bigl|m^2 - {(m')}^2\bigr| \,,
\label{eq:resonance-x}
\end{align}
and it follows from \eqref{eq:3+1Ahat-int-components}
that the mode mixing growth rate is 
\begin{align}
\tfrac{d}{d\tau}|\Ahat_{\text{res}}| 
& \approx 
\tfrac12 \pi m m' d_x\lambda L_x^{-3}
\,. 
\label{eq:Ahatresabsdot-x}
\end{align}
The lowest resonance
occurs for $m=1$ and $m'=2$. 
Similar formulas ensue for the $y$-resonance, and for circular motion 
both resonances are present. 

As the experimental setup, we first trap one or more quanta in the
cavity, in modes whose momenta are aligned close to the $z$-direction.
After a period of linear or circular oscillation perpendicular to the
$z$-direction, a measurement on the quantum state of the cavity is
performed, by suitable observations of quanta that are allowed to
escape.  We assume that the resonance mode mixing dominates any
effects due to the initial trapping and the final releasing of the
quanta.

A careful choice of the cavity geometry would need to be 
considered in order to guarantee the success of an experiment. 
A~particular concern would be the mechanical stability of the cavity itself. 
There are options for creating mechanically very robust 
cavities based on a monolithic geometry. Examples could 
be a Bragg grating cavity in an optical fibre. 
These are mechanically very robust and are a very well developed technology. 
Another example, closer to the parameters specified below, 
would be a monolithic Fabry-Perot cavity or etalon filter cavity. 
Such cavities can be made with extremely high finesse and are 
made out of a single solid block of material and hence inherit 
the robustness of the material itself (typically, glass). 

We choose $\lambda = 600\,$nm and 
$L_x = L_y= 1\,$cm. 
The lowest resonance angular frequency is then 
$\omega_c \approx 
1.4 \times 10^{-2} \, {\text{m}}^{-1}
\approx 
4.2 \times 10^{6} \, {\text{s}}^{-1}$, 
corresponding to an oscillation frequency
$0.7\,$MHz. 

For linear oscillation, we choose the amplitude $d_x = 1\,\mu$m, 
which may be achievable by using ultrasound to accelerate the cavity. 
From \eqref{eq:Ahatresabsdot-x} we then have 
$\tfrac{d}{d\tau}|\Ahat_{\text{res}}| \approx 
6 \times 10^{2} \, {\text{s}}^{-1}$, 
so that the mode mixing coefficient grows to 
order unity within a millisecond. 
For the squeezed states of Section~\ref{subsec:resonances}, 
the negativity \eqref{eq:firstorderneg} grows to 
order unity at the same timescale provided the 
squeezing parameter $s$ is not much less than unity. 
Storing the quantum in the cavity for a millisecond could be challenging 
although recent achievements indicate that it may be feasible~\cite{ultra-fabry-perot}.

For circular motion, 
we choose the amplitude $d_x=d_y=1$mm. 
At the threshold angular velocity 
$\omega_c \approx 4.2 \times 10^{6} \, {\text{s}}^{-1} \approx 4\times 10^7\,$rpm, 
the mode mixing coefficient then grows to order unity within a nanosecond, 
and for squeezed states similarly for the negativity \eqref{eq:firstorderneg} 
provided the squeezing parameter $s$ is not much less than unity. 
The threshold angular velocity exceeds the angular velocity of 
medical ultracentrifuges by a factor 
of 200~\cite{thermocientific}, 
but this gap could possibly be bridged by a specifically 
designed system of sub-centimetre scale. 
We note that the centripetal acceleration at the threshold angular velocity equals 
$1.5 \times 10^6 \, {\text{m}} {\text{s}}^{-2}$, 
which is already reached in ultracentrifuges that combine 
a smaller angular velocity with a larger radius~\cite{thermocientific}. 

As $\omega_c$ in these scenarios 
is much below the particle creation 
resonance~\eqref{eq:3+1resonance-Bhat}, 
particle creation in the cavity is not cumulative in the 
duration of the oscillation and is highly sensitive to the 
manner in which the acceleration is switched on and off. 
While this is a consequence of the idealised, 
fully confining character of our cavity, 
we may obtain an upper limit for the predicted 
particle creation by noting that in the extreme case of 
sharp switch-on and switch-off 
$\Bhat_{mnp,m'np}$ \eqref{eq:3+1Bhat-int-components} 
has the order of magnitude 
\begin{align}
\frac{\pi^2 m m' \bigl(1 - {(-1)}^{m+m'}\bigr) |a_x|L_x} 
{L_x^4\left(\omega_{mnp} + \omega_{m'np}\right)^3
\sqrt{\omega_{mnp} \omega_{m'np}}}
\ . 
\label{eq:3+1Betaorder}
\end{align}
The number of particles created in a mode with fixed $m$, $n$ and $p$, 
each near their lowest value $1$, can hence be given an upper bound by summing 
the square of \eqref{eq:3+1Betaorder} over~$m'$. The result is 
a purely numerical factor times ${(a_x L_x)}^2$, 
which is of order $10^{-24}$ for our linear oscillation 
figures and of order $10^{-18}$ for our circular motion figures. 
At the mode mixing resonance, the mixing hence overwhelmingly 
dominates over any particle creation effects. 
This is consistent with the usual estimates of 
$10^{-9}$ photons created per second \cite{Reynaud1} 
for experimentally less idealised cavities.

\section{Conclusions}

We have quantised a scalar field in a rectangular cavity that is
accelerated arbitrarily in $(3+1)$-dimensional Minkowski spacetime, in
the limit of small accelerations but arbitrary velocities and travel
times. The Bogoliubov coefficients were expressed as explicit
quadratures. For linear or circular periodic motions, we identified a
configuration in which the mode mixing resonance frequency is
significantly below the frequencies of the cavity modes. 

Our scalar field analysis adapts in a straightforward way to a Maxwell
field with perfect conductor boundary
conditions~\cite{FLL:manyspins}. The mode mixing effects appear hence
to be within the reach of a desktop experiment with photons,
achievable with current technology in its mechanical aspects, if
perhaps not yet in the storage capabilities required of a mechanically
oscillating optical cavity.


We anticipate that the particle creation and mode mixing effects are 
not qualitatively sensitive to the detailed shape of the cavity, 
and this freedom could be utilised in the development of a concrete 
laboratory implementation. The experimental prospects could be further 
improved by filling the cavity with a medium that slows 
light down~\cite{laupretre-slowlight}. 
A~laboratory implementation would also need to develop 
an experimental protocol for measuring the field within the cavity, 
and the data analysis 
would need to account for any experimental imperfections. 
A~full detailed evaluation of these experimental issues 
would need to be carried out 
case by case for any proposed concrete implementation, 
but the frequency and lifetime estimates given in this paper 
do suggest the mode mixing effect to be at the threshold of current technology. 


We underline that our experimental scenario does not involve significant particle creation. 
Nevertheless, it involves significant mode mixing. 
This mixing acts as a beam splitter quantum gate, creating or degrading 
entanglement in situations where particles are initially present. 
Finally, it is also worth underlining that although we have discussed the 
specific case of a mechanically oscillating cavity, the 
low-frequency resonance can be found whenever the quanta can be made highly transverse to the
acceleration, and may therefore 
be similarly adopted to perform quantum gate operations also in other analogue systems, based 
e.g.\ on SQUID mirrors \cite{wilson-etal} or nonlinear optics \cite{lambrecht,carusotto} 
that have been proposed to date.  
We anticipate that observations of
entanglement will generally provide opportunities for experimental
verification of both particle creation and mode mixing effects that
are complementary to observations of fluxes or particle numbers~\cite{wilson-etal}.


\begin{acknowledgments}

We thank Nico Giulini and Bill Unruh for invaluable discussions. 
We also thank 
Iacopo Carusotto, 
Fay Dowker, 
Gary Gibbons, 
Chris Fewster,
Nico Friis,  
Vladimir Man'ko,
Carlos Sab\'in, 
Ralf Sch\"utzhold 
and 
Matt Visser for helpful comments. 
J.L. thanks 
Gabor Kunstatter for hospitality at the 
University of Winnipeg and 
the organisers of the 
``Bits, Branes, Black Holes'' 
programme for hospitality at the 
Kavli Institute for Theoretical Physics,
University of California at Santa Barbara, 
supported in part by 
the National 
Science Foundation under Grant No.\ NSF PHY11-25915. 
D.F. acknowledges financial support from the 
Engineering and Physical Sciences Research Council EPSRC,
Grant EP/J00443X/1 and from the European Research Council 
under the European Union's Seventh Framework
Programme (FP/2007-2013) / ERC Grant Agreement n.306559.
I.F. acknowledges financial support from EPSRC [CAF Grant EP/G00496X/2].
J.L. acknowledges financial support from STFC [Theory Consolidated
Grant ST/J000388/1].

\end{acknowledgments}

\end{document}